\documentclass[11pt]{article}
\usepackage{moriond,epsfig}

\def\MyPhoto{\pdfimage width 35mm {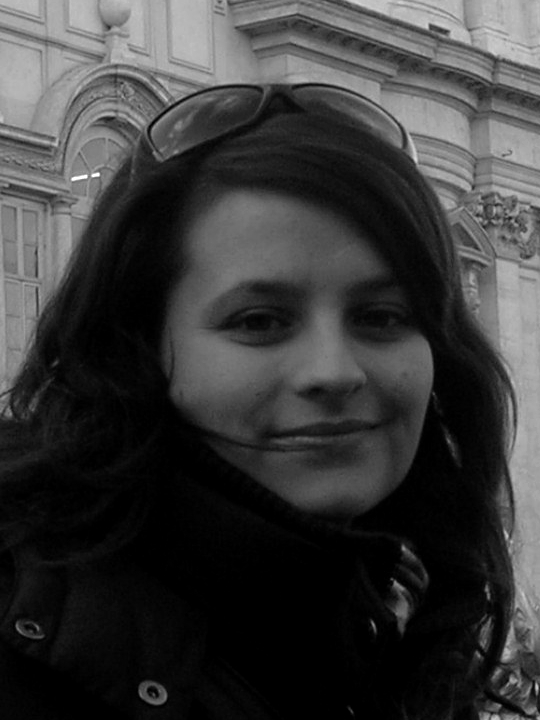}}

\def\Journal#1#2#3#4{{#1} {\bf #2}, #3 (#4)}

\def\NIMA{{\em Nucl. Instrum. Methods} A}

\def\PRD{{\em Phys. Rev.} D}

\def\be{\begin{equation}}
\def\ee{\end{equation}}
\def\bea{\begin{eqnarray}}
\def\eea{\end{eqnarray}}

\begin{document}
\vspace*{4cm}
\title{STATUS AND FIRST RESULTS OF THE ANTARES NEUTRINO TELESCOPE}

\author{ GIADA CARMINATI \\ on behalf of the ANTARES Collaboration}

\address{Dipartimento di Fisica, Universit\`a e Sezione INFN di Bologna, \\
Viale Berti Pichat 6/2, 40127 Bologna, Italy}

\maketitle\abstracts{ The ANTARES (Astronomy with a Neutrino
Telescope and Abyss environmental RESearch) Collaboration
constructed and deployed the world's largest operational underwater
neutrino telescope, optimised for the detection of Cherenkov light
produced by neutrino-induced muons. The detector has an effective
area of about 0.1 km$^2$ and it is a first step towards a kilometric
scale detector. The detector consists of a three-dimensional array
of 884 photomultiplier tubes, arranged in 12 lines anchored at a
depth of 2475 m in the Mediterranean Sea, 40~km offshore from Toulon
(France). An additional instrumented line is used for environmental
monitoring and for neutrino acoustic detection R\&D. ANTARES is
taking data with its full twelve line configuration since May 2008
and had been also doing so for more than a year before a five and
ten line setups. First results obtained with the 5 line setup are
presented. }

\vspace*{-7mm}
\section{Introduction}

The production mechanism (i.e. the cosmic accelerators) of high
energy cosmic rays remains one of the open problem in astroparticle
physics. The detection of high energy cosmic neutrinos would help to
find some answers. Pursuing this aim, the ANTARES Collaboration
completed in 2008, and has been operating since, a neutrino
telescope off the Southern French coast, at a depth of 2475 m under
the sea level.

ANTARES can be seen as a fixed target experiment: a cosmic muon
neutrino, produced in a cosmic source (Supernova remnant, AGN, GRB,
$\ldots$) arriving mainly from the hemisphere opposite to the
detector location, crosses the Earth and interacts by a charged
current process with a nucleon of the medium surrounding the
telescope and induces a muon. Above a few TeV the neutrino-induced
upward-going muon is (almost) collinear with the incident neutrino
and can travel up to 10 km in the rock reaching the sea water. The
Cherenkov light emitted by the muon with an angle $\theta_C \simeq
42.2^\circ$ is detected by the three-dimensional array of 884
photomultiplier tubes (PMTs) that comprises the ANTARES detector.

However, the most abundant signal is due to high energy
downward-going muons remaining from the extensive air showers
produced in interactions between primary cosmic rays and atmospheric
nuclei. Although the shielding effect of the sea reduces their flux,
at the ANTARES site the atmospheric muon flux is about six orders of
magnitude larger than the atmospheric neutrino flux. These
atmospheric muons represent a dangerous background for track
reconstruction as their Cherenkov light can mimic fake upward-going
tracks. On the other hand, they are a useful tool to test offline
analysis software, to check the understanding of the detector and to
estimate systematic uncertainties. The other main background is due
to the atmospheric neutrinos that have the same signature as the
cosmic signal ANTARES awaits for: un upward-going muon, but with
different the energy spectrum.

\begin{figure}[!h]
\centering
\includegraphics[scale=0.7]{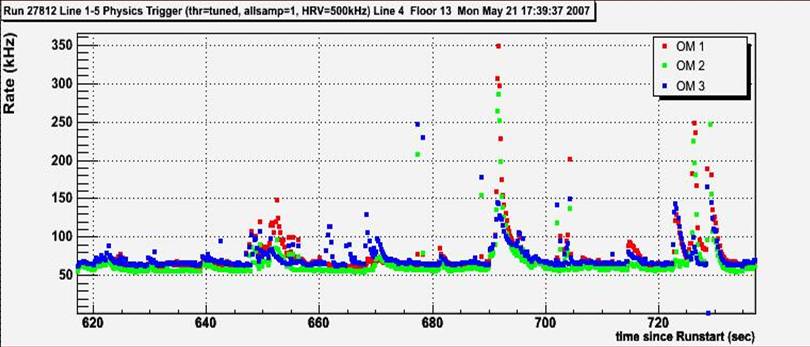}
\caption{ Optical background rate measured by 3 PMTs of a storey in
a time window of 2 minutes (May 2007).} \label{fig:rate}
\end{figure}

The PMTs also detect the light generated by $^{40}K$ decays in the
sea water and the bioluminescence emitted by marine organisms. These
two light sources give a continuous background, which varies between
60 and 100 kHz. Peaks of biological activity can occasionally
increase the counting rate up to the order of several MHz. Figure
\ref{fig:rate} shows the typical background behaviour during data
taking. Fortunately the hits produced by $^{40}K$ decays and
bioluminescence are mostly uncorrelated and can be easily discarded
by the trigger algorithm.

\vspace*{-6mm}
\section{The ANTARES detector}

The detector consists of 12 strings anchored on the seabed and
pulled up by a buoy and by the buoyancy of the instrument
containers. A string is 480 m long and is composed of 25 storeys. A
storey includes 3 optical modules, each made of a 17"
pressure-resistant glass sphere housing a 10" PMT~\cite{PMT},
looking downwards at 45$^\circ$ from the vertical. Due to the
flexibility of the lines, a redundant positioning system is
incorporated on each line. It includes 25 tiltmeters and compasses
measuring the roll, the pitch and heading angles of the storey, and
5 hydrophones by which the positioning is determined using acoustic
triangulation between fixed emitters on the sea floors. This
information allows the shape of the lines to be reconstructed and
the position accuracy to be better than 10~cm.

Each storey is also equipped with a titanium electronics container,
where the PMT signal is digitized by an analogue ring sampler (ARS),
a custom-built microchip which measures the charge and the arrival
time of the pulse with a precision better than 1 ns. Each readout
channel includes two ARS working in a token ring mode to reduce the
dead time. The signal is then treated by a data acquisition card and
sent through optical fibres to a main control module and then to the
bottom of the line where it is multiplexed by DWDM. The fibres of a
line pass through the junction box, where the lines are connected to
a 40 kilometer electro-optical submarine cable going to the shore.

The large bandwidth of the DAQ system allows the transmission of all
recorded PMT signals, above a given threshold, to shore. At the
shore station, a dedicated computer farm runs various trigger
softwares~\cite{trigger} to select the photon hits on the PMTs of
the interesting physics events from the data recorded by the whole
detector. The photon hits filtered by the trigger make up a
so-called event. All the events are stored on disks for further
track reconstruction and analysis. The knowledge of the hit
position, arrival time and charge allows the reconstruction of the
muon trajectory thus giving information on the parent neutrino.

The first ANTARES detection line was connected on March 2006. Five
lines of the detector have been operated since the end of January
2007, followed by 5 additional lines in December of the same year.
The detector has been completed on May the 30$^{\mathrm{th}}$, 2008,
with the connection of the last two lines.

\vspace*{-2mm}
\section{The first physics analysis}
The analysis of atmospheric muons in the data taken in 2006 with the
first detection line has been reported in~\cite{line1}. In 2007,
ANTARES has taken data with the 5 line setup for 314 days. The duty
cycle of the data taking in this period has been of about 80\%. As a
first step, the analysis was restricted to runs with a very low
biological activity, i.e. the presence of bioluminescence peaks for
less than 40\% of a run duration. This corresponds to an equivalent
data taking of 140 days, in which 14.5 million events have
triggered.

\vspace*{-3mm}
\subsection{Atmospheric muons}

For the study of atmospheric muons, a more strict selection was
applied to the data set: only runs where the presence of
bioluminescence peaks was less than 20\% of a run duration are used.
The analysis is also limited to the period from June to December
2007 corresponding to 76 active days of data taking, during which
about 7.5 million events have triggered.

Monte Carlo (MC) simulations of the detector response have been used
to understand the data. Several MC samples have been produced, using
different input parameter sets. ANTARES uses two different
approaches to simulate the atmospheric muon flux: a parameterized
description of underwater muon flux and a full MC simulation. The
former is based on the MUPAGE code~\cite{mupage} which generates
events directly on the active volume surrounding the detector. The
full MC simulation uses the CORSIKA software~\cite{corsika} and the
hadronic interaction model QGSJET.01c~\cite{qgs}, to generate a
large number of air showers induced by primary cosmic ray nuclei,
and the MUSIC code~\cite{music} to propagate muons through sea water
until the active volume surrounding the detector. Two different
compositions of the primary cosmic ray flux are taken into account:
the NSU model~\cite{nsu} and a simplified version of the H\"orandel
model~\cite{horandel}.

\begin{figure}[!h]
\centering
\includegraphics[scale=0.6]{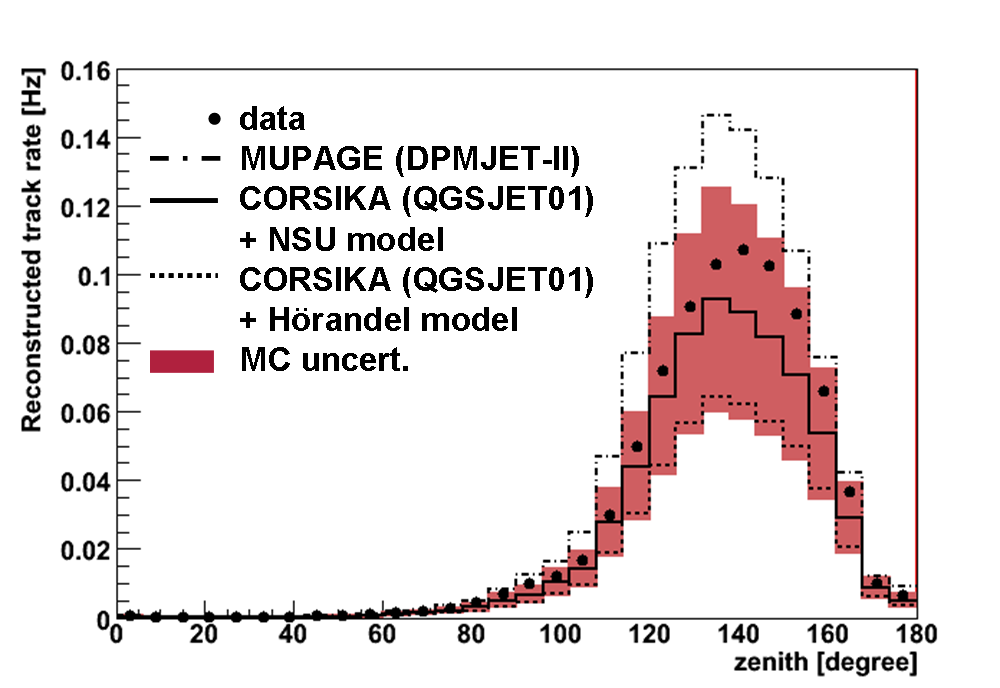}
\caption{Zenith distribution of reconstructed tracks (180$^\circ$  =
downward-going). Black points represent data and the statistical
errors are within the point size. Lines refer to MC expectations:
solid line = full simulation with CORSIKA, QGSJET01 and NSU model;
dotted line = full simulation with CORSIKA, QGSJET01 and
H\"{o}randel model; dashed-dotted line = MUPAGE parameterisation,
with DPMJET-II for hadronic interaction. The shadowed band
represents the systematic errors, starting from the NSU model, due
to environmental and geometrical parameters.} \label{fig:atm-muons}
\end{figure}

The effect of environmental and geometrical parameter uncertainties
on reconstructed track rate is also studied~\cite{annarita}.
Presently, a value of about $\pm$35\% can be considered as a first
evaluation of the systematic errors and is shown, as a shadowed
band, in Figure~\ref{fig:atm-muons}. New measurements are in
progress to reduce uncertainties on the parameters used as input for
MC simulations.

\vspace*{-3mm}
\subsection{Neutrino candidates}

\begin{figure}[!b]
\begin{minipage}[t]{0.47\linewidth}
\includegraphics[scale=0.43]{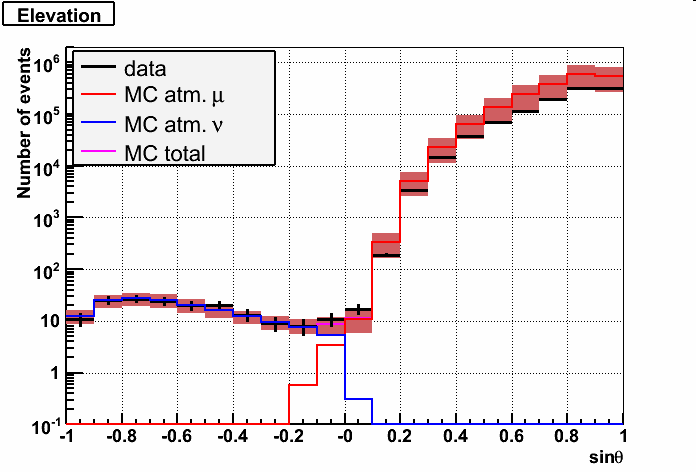}
\centering \caption{Number of reconstructed events vs the sine of
the elevation ($\sin(\theta)=1$: downward-going). The black line
represents data, the red and blue lines represent, respectively, the
MC atmospheric muons and neutrinos, the sum being represented by the
pink line. }\label{fig:neutrino_l05}
\end{minipage}
\hspace{0.5cm}
\begin{minipage}[t]{0.47\linewidth}
\includegraphics[scale=0.5]{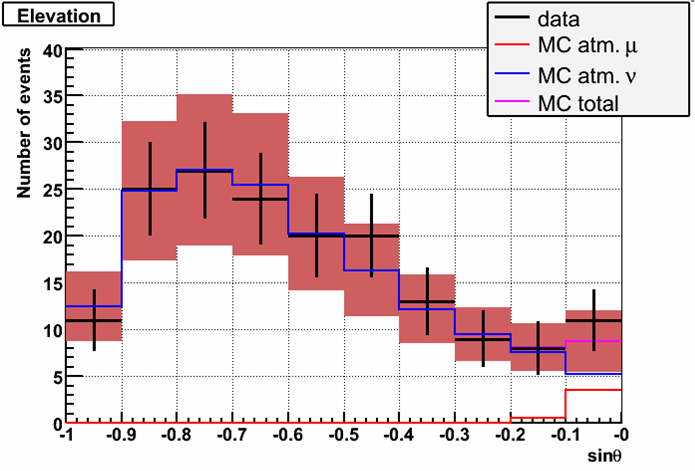}
\centering \caption{Zoom of the number of reconstructed events as a
function of the sine of the elevation, focusing on the neutrino
candidates.} \label{fig:neutrino_l05_cut}
\end{minipage}
\end{figure}

The 5 line data have been analysed in order to select the neutrino
candidates. Figure~\ref{fig:neutrino_l05} shows the result of one of
the ongoing studies: the number of reconstructed events as a
function of the elevation is plotted. For the simulation of
atmospheric muons (red line) the full MC simulation with the NSU
model was used. The Bartol model~\cite{bartol} has been used to
simulate the atmospheric neutrino flux (blue line). The negative
$\sin{\theta}$ events are the upward-going events, the neutrino
candidates. Figure~\ref{fig:neutrino_l05_cut} is a zoom. The events
are reconstructed by an on-line reconstruction code, after the
application of some quality cuts. The selection yields 168 neutrino
candidates from the data, while 161$\pm$32(theo)$\pm$20(syst) were
expected from the MC simulations.

A good agreement between data and simulation is also found with
another independent atmospheric neutrino analysis, based on a
different reconstruction algorithm~\cite{aart}. These results prove
the correct detector response to a known neutrino source such as the
atmospheric neutrinos.

\vspace*{-3mm}
\subsection{Point-like neutrino sources}

\begin{figure}[!h]
\begin{minipage}[t]{0.47\linewidth}
\includegraphics[scale=0.5]{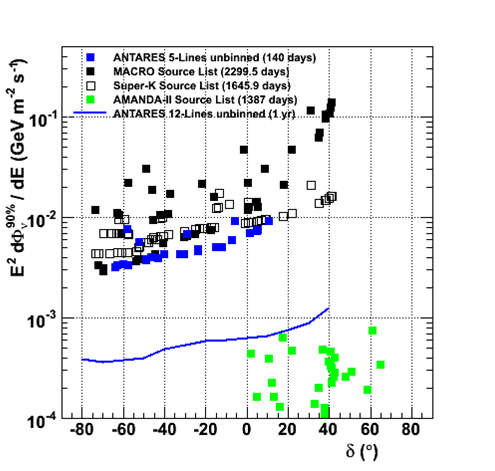}
\centering \caption{Neutrino flux upper limit for selected sources
(blue points) for a $E^{-2}$ spectrum as a function of declination
using the 5 line data. Also shown are the predicted sensitivity for
1 year of the 12 line setup (blue solid line) and upper limits for
selected sources by Super-Kamiokande (empty black points), MACRO
(full black points) and AMANDA-II (green
points).}\label{fig:point-like_source}
\end{minipage}
\hspace{0.5cm}
\begin{minipage}[t]{0.47\linewidth}
\includegraphics[scale=0.6]{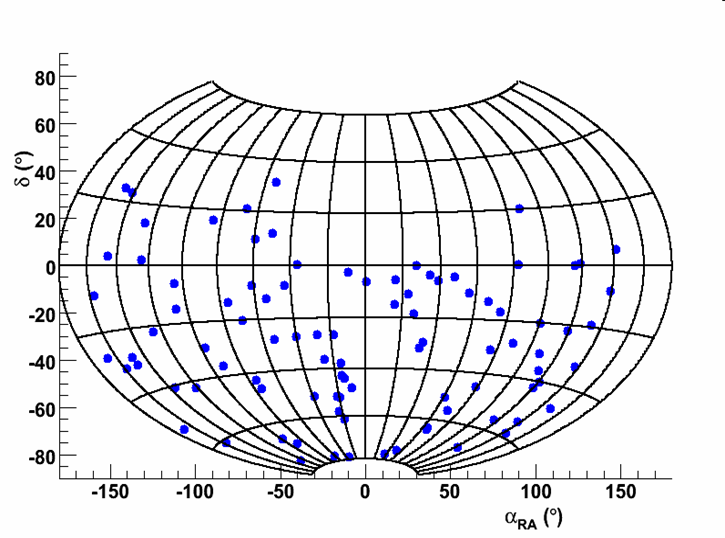}
\centering \caption{Sky map in equatorial coordinates of 94 neutrino
events detected with the 5 line setup.}
\label{fig:candidate_pointlike}
\end{minipage}
\end{figure}

The 5 line data have been used to search for point like sources in
the Southern hemisphere. Applying optimised cuts, the angular
resolution is below 0.5$^\circ$ above 10 TeV. No cosmic neutrino
neutrino signal is found at 90\% confidence level. The differential
flux upper limit for a list of selected sources is shown in
Figure~\ref{fig:point-like_source} as a function of the declination
of the sources (blue points) and is compared with the results
obtained by other experiments. It is interesting to note that the
sensitivity for 140 active days on the 5 line data is already better
than those of the many year detection of SuperKamiokande~\cite{SK}
(empty black point) and MACRO~\cite{macro} (full black point). This
is because ANTARES is more efficient at high energies, which is the
critical region for this kind of searches (since we assume that the
signal emits as $E^{-2}$). For the opposite hemisphere the results
of AMANDA-II~\cite{amanda} are shown (green points). The predicted
sensitivity for 1 year of the full ANTARES configuration (solid blue
line) is also shown. Although no signal is found,
Figure~\ref{fig:candidate_pointlike} shows the sky map of the 94
neutrino events considered in this analysis.

\vspace*{-3mm}
\subsection{Dark matter}

\begin{figure}[!h]
\begin{minipage}[t]{0.47\linewidth}
\includegraphics[scale=0.5]{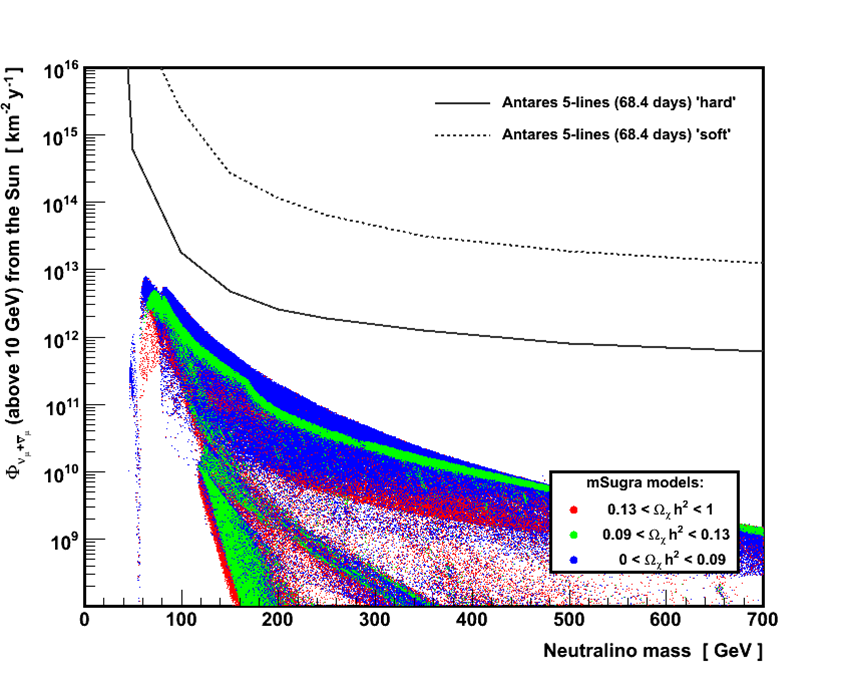}
\centering \caption{Upper limit on the total $\nu_\mu$ +
$\overline{\nu_\mu}$ flux from neutralino annihilation in the Sun,
after analysing 5 line data. Each coloured point corresponds to a
supersymmetric model.}\label{fig:DM-l05}
\end{minipage}
\hspace{0.5cm}
\begin{minipage}[t]{0.47\linewidth}
\includegraphics[scale=0.53]{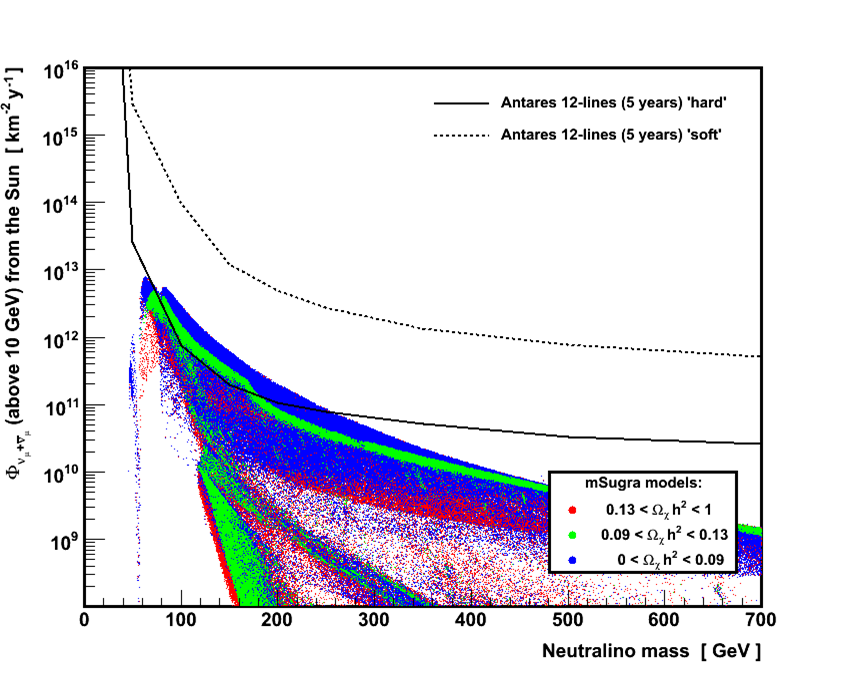}
\centering \caption{Prediction of the upper limit on the total
$\nu_\mu$ + $\overline{\nu_\mu}$ flux from neutralino annihilation
in the Sun for the full ANTARES configuration after 5 years of data
taking. Each coloured point corresponds to a supersymmetric model }
\label{fig:DM-l12}
\end{minipage}
\end{figure}

Dark matter searches are also possible by looking for an excess of
neutrinos from celestial bodies like the Sun or the Galactic Center.
Neutrinos with energy below a TeV could be produced by the
annihilation of supersymmetric dark matter particles like the
neutralino, that would become gravitationally trapped in these
bodies. For neutralinos in the Sun, the analysis of the 5 line data
is reduced to about 70 active days, because only data when the Sun
is under the horizon are analysed. Figure~\ref{fig:DM-l05} shows the
upper limit on the total flux from neutralino annihilation in the
Sun with the 5 line data as function of the neutralino mass. Each
point corresponds to a supersymmetric model and different
observational constraints from WMAP~\cite{wmap}. Two kinds of
annihilation (hard, into $W$ vector bosons and soft, into $b\bar{b}$
quarks) have been studied. Figure~\ref{fig:DM-l12} shows the
prediction for 5 years of data taken with the 12 line configuration.

\vspace*{-2mm}
\section{Summary and outlook}
ANTARES is now the largest neutrino telescope in the Northern
hemisphere. The analysis of the data taken in 2007 with the first 5
lines of the telescope has been performed. Concerning the
downward-going muon flux, agreement has been found between the data
and some of the Monte Carlo simulation models, still under study. A
sample of 168 neutrino candidate events has been selected in 140
days of active time. A search for high energy neutrinos produced
with an $E^{-2}$ energy spectrum has also been performed on the
basis of a potential source list, yielding to an upper limit better
than the ones obtained after many years by Super-Kamiokande and
MACRO. A first limit to the neutrino flux from supersymmetric dark
matter annihilation in the core of the Sun has been set.

ANTARES is also meant as a first step towards a larger detector in
the Mediterranean Sea. The ANTARES Collaboration is involved in the
KM3NeT Consortium~\cite{km3net}, accepting the challenge of
producing a cubic kilometer neutrino telescope.


\vspace*{-2mm}
\section*{References}
\begin{thebibliography}{99}
\bibitem{PMT} The ANTARES Collaboration, \Journal{\NIMA}{484}{369}{2002};
\Journal{\NIMA}{555}{132}{2005}.

\bibitem{trigger} The ANTARES Collaboration, \Journal{\NIMA}{570}{107}{2007}.

\bibitem{line1} The ANTARES Collaboration, \Journal{\em{Astrop.
Phys.}}{31}{277}{2009}.

\bibitem{mupage} G. Carminati {\it et al}, \Journal{\em{Comput. Phys.
Commun.}}{179}{915}{2008}.

\bibitem{corsika} D. Heck {\it et al}, Report FZKA6019 (1998),
Forshungszentrum Karlsruhe; D. Heck and J. Knapp, Report FZKA6097
(1998), Forshungszentrum Karlsruhe.

\bibitem{qgs} N.N. Kalmynov and S.S. Ostapchenko, \Journal{\em{Yad.
Fiz.}}{56}{105}{1993}; \Journal{\em{Phys. At.
Nucl.}}{56}{346}{1993}; N.N. Kalmynov, S.S. Ostapchenko and A.I.
Pavlov, \Journal{\em{Bull. Russ. Acad. Sci.
(Physics)}}{58}{1966}{1994}.

\bibitem{music} P. Antonioli {\it et al}, \Journal{\em{Astrop.
Phys.}}{7}{357}{1997}.

\bibitem{nsu} S.I. Nikolsky, J.N. Stamenov and S.Z. Ushev,
\Journal{\em{Sov. Phys. JETP}}{60}{10}{1984}; \Journal{\em{Zh. Eksp.
Teor. Fiz.}}{87}{18}{1984}.

\bibitem{horandel} J.R. H\"orandel, \Journal{\em{Astrop.
Phys.}}{19}{193}{2003}.

\bibitem{annarita} A. Margiotta, \Journal{\NIMA}{602}{76}{2008}.

\bibitem{bartol} G. Barr, T.K. Gaisser and T. Stanev,
\Journal{\PRD}{39}{3532}{1989}.

\bibitem{aart} A. Heijboer, PhD Thesis, http://antares.in2p3.fr
(2005).

\bibitem{SK} Super-Kamiokande Collaboration, \Journal{\em{Astrophys.
J.}}{652}{198}{2006}.

\bibitem{macro} MACRO Collaboration, \Journal{\em{Astrophys.
J.}}{546}{1038}{2001}.

\bibitem{amanda} IceCube Collaboration, \Journal{\PRD}{79}{062001}{2009}.

\bibitem{wmap} WMAP Collaboration, \Journal{\em{Astrophys.
J.}}{148}{175}{2003}.

\bibitem{km3net} KM3NeT Consortium: http://www.km3net.org

\end{thebibliography}
\end{document}